\documentclass[12pt]{article}
\usepackage{epsfig}
\begin{document}
\title{A model of information filtration
       by comparison of randomly chosen sources}
\author{
I.S.Manida,  Yu.M.Pis'mak\\
\\
\small
{\it
Department
of Theoretical Physics
State University of
Saint-Petersburg}\\
\small
{\it
Ul'yanovskaya 1, Petrodvoretz,
195904 Saint-Petersburg, Russia}\\
}
\date{}
\maketitle
\abstract
{
We study a simple model of the stochastic information filtering,
in a randomly organized information system. For simplest
versions of the model it appears to be possible to describe the filtering
dynamics in terms of the master equations. Exact analytical results
for these equations and results of numerical investigation of the
dynamical features of the filter are presented.
}

\section{Introduction}
The main task  of many  modern information technologies  is
to  increase  the velocity  of information processing.
The important theoretical problem in this region is to find
the basic principles of  efficient  organization of
information  flows.   One of the possible approaches for its
solution is developed in the framework of  mathematical modeling
of universal  self-organization mechanisms in complex dynamical systems.

Often in this investigation, it appears to be useful to
have  some  physical picture of phenomena under consideration.
Usually, one can imagine the information system, as  a net of
the connected information sources. Its physical role is to
realize interaction between suppliers and users of information.
Although this interaction forms laws of
the system evolution,  in many situations it  can be presented
effectively as some  specific of elementary dynamical rules
for the sources in the model of information net.

Recently a model of this type was  proposed by
A. Capocci, F.Slanina 
and Y.-C. Zhnag  \cite{Zhang}. It describes  ranking
and filtration
of information,
being widespread  procedures  of  information processing.
The investigated in \cite{Zhang} model is a 1-dimensional
stochastic dynamical system with nearest neighbor
interaction.

The structure of many information system
in reality is very complex, and
frequently, it can be better presented by a random graph than
by 1-dimensional chain.
Experience in studies of  physical phenomena shows that
normally the interaction structure influences  essentially
on the system behavior. Therefore it seems to be
interesting to investigate the random neighbor version of
the  proposed in \cite{Zhang} model
\footnote {The statement of this problem  was formed
in discussion with Y.-C. Znahg of the
presented in \cite{Zhang}
principles of information filtering. }.  We present in this paper
the numerical and analytical results obtained for
information filter dynamics  of such a kind.

\section{Formulation of model}
We will consider a version of the filter dynamics without a fixed
interaction topology. The simplest modification of the model
\cite{Zhang} could be
formulated in the following way. There are n elements with
characteristics called qualities and being a number from interval
\lbrack 0,1\rbrack . The initial state of the system is chosen at
random. The state at the time point \ $t+1$ is obtained as follows.
One chooses at time point $~t~$ some 2 random
elements and at the time point $t+1~\ $the qualities of both
elements will be equal to the  same number $Q(t+1)$. We consider
two versions of the model - model A (MA) and model B (MB). In the
MA the elements can be chosen arbitrarily. In the MB  the quality
of the chosen elements must be different, and when there are not
different elements its state does not change. If the qualities of
elements chosen at time $t~$ are  $q$ and $p$ then $Q(t+1)=q$ with
probability $\frac{q}{p+q}=\mu~$and $Q(t+1)=p$ with probability
$\frac{p}{p+q}=\lambda~$. If the qualities of the chosen elements
were the same, it does not change. Thus, in both models the state
of the system becomes stable if all of its elements are of the
same quality.
   The dynamics of proposed models can be investigated by exact analytical
methods. We demonstrate it for the initial condition of a special case.
We suppose that in the initial state $l$elements are of the quality $p$
and $m=n-l$ of  the quality $q$.  This dynamical situation arouses in the
general case  until the system reach the stable state. The qualitative
description of dynamics of such kind can be obtained with the help of
master equations.

\section{Numerical results}

We have chosen the MA model for numerical analysis as it is more
complicated for analytical studies. Numerical experiment have been performed
with the help of a distributed C program running on Solaris cluster,
calculating all possible combinations of model variations, initial
distributions, and number of elements;
with following analysis of obtained results in
Mathematica to detect model behaviour patterns.

The main characteristic of MA dynamics appears to be the scale
invariance in respect to
the number of elements $N$, which is independent from the model variations
and initial conditions.
We studied the system's average value of elements
quality
$$
A(t,N) \equiv  <Q(t,N)>
$$
where $Q(t,N)$ denotes the mean value of the system's element quality at the
time point $t$ and $<...>$ means averaging over ensembles of
system evolution.
Our results show that we can write $A(t,N)$ in the following form:
\begin{eqnarray}
A(t,N)= A_{inv}\left(\frac{t}{N}\right)
\label{num1}
\end{eqnarray}
if the average values $A(0,N)$  given by initial conditions are
independent from $N$.
Dynamics of $A(t,N)$ for the MA model with two types of initial states
are shown on the fig. 1.
We considered the systems with initial states
equidistributed
on the interval [0,1]
($A(0,N)=0,5$) and initial states with 10\%
elements of the
quality 0,9 and 90\% of elements of the quality 0,1
($A(0,N)=0,1\cdot 0,9 + 0,9\cdot 01=0,18$).
For large systems with $N>100$ we obtained a close correspondence
between numerical results for $A(t,N)$ and (\ref{num1}).
For equidistributed
initial condition the correspondence is shown on the fig. 2. For this type
of initial condition we have observed
\begin{eqnarray}
A_{inv}\left(\frac{t}{N}\right)=1-\left(2+\frac{t}{N}\right)^{-1}.
\label{num2}
\end{eqnarray}
For small $N$ there is a certain deviation of experimental curve $A(t,N)$
from (\ref{num1}), (see fig. 3), which is caused by strong effect of random
fluctuations of systems with small number of elements.
We have also calculated the average deviation from the average value
$$
D(t,N)=\sqrt{<(Q(t,N)-A(t,N))^2>}.
$$
The typical curve for $D(t,N)$
is presented on the fig. 4.

   For initial state equidistributed on interval [0,1] we investigated
the averaged value of number ${\cal N}(t,N)$ of different elements in the system
The obtained curve for ${\cal N}(t,N)/N$ seems to be very well approximated by
the function $S(t,N)=N/(t+N)$ (see fig. 5).
   We modified dynamics of the MA by adding Darwin selection to
the system. By using the model of natural selection mechanism
proposed in \cite{Bak}, it was presented  as follows: at each time step
the state of the system is changed by the MA rules and after that
the element with lowest quality in the system is replaced by the one with
an arbitrarily chosen quality in the interval [0,1].

The influence of
the selection mechanism on the system dynamics can be analyzed by
comparison of the process in usual MA and MA with Darwin selection
for the same initial conditions as it is presented on the fig. 6 for a system with elements having one of
two quality values and different initial distributions of "bad" (quality 0.9) and "good" (quality 0.1) elements.
We see that the Darwin selection speeds up the filtration process
for $A(0,N)<0,5$ and slows it down for $A(0,N)>0,5$.

\section{Master equations}

We denote the state of the system having $k$ elements of the quality
$p$ and $n-k$ elements of the quality $q$ at the time point $t$
as $\{k;t\}$. Let $P_{k}(t)~$ be the probability of the state
$\{k;t\}$. If at the time point $t+1~$\ the system state is
$\{k,t+1\}$ then it was in one of the states $\{k-1;t\}$,
$\{k;t\}$  or $\{k+1;t\}$ at the previous moment $t$.
The probability $P_{2,0}(k)~$
to choose two elements of quality $p$ in the state $\{k;t\}$
is
\[
P_{2,0}(k)=\frac{k(k-1)}{n(n-1)}.
\]
The probability $P_{0,2}(k)~$to choose two elements of quality $q$ is
\[
P_{0,2}(k)=\frac{(n-k)(n-k-1)}{n(n-1)}.
\]
The probability $P_{1,1}(k)$ to choose one element of quality $p$ and
one of quality $q$ is
\[
P_{1,1}(k)=\frac{2k(n-k)}{n(n-1)}.
\]

For $k\neq 0$, \ $k\neq n$ , the probabilities  $P(\{k;t\},
\{k';t+1\})$  of transition
$\{k;t\}\longmapsto \{k';t+1\}$ in the MA  can be presented as
$$
P(\{k;t\},\{k;t+1\})=P_{2,0}(k)+P_{0,2}(k),
$$
$$
P(\{k+1;t\},\{k;t+1\})=P_{1,1}(k+1)\mu ,
$$
$$
P(\{k-1;t\},\{k;t+1\})
=P_{1,1}(k-1)\lambda
$$
where $\lambda = p/(p+q)$, $\mu = q/(p+q)$. For  $k=0$ or $k=n$,
the state $\{k;t\}$ of the system is stable, hence
\[
P(\{0;t\},\{0;t+1\})=P(\{n;t\},\{n;t+1\})= 1.
\]
Thus, we can write the master equation for probability $P_{k}(t)$ as follows:
\[
P_{k}(t+1)= \lambda P_{k-1}(t)P_{1,1}(k-1)(1-\delta _{k0})+
\]
\[
+ \mu P_{k+1}(t)P_{1,1}(k+1)(1-\delta _{kn})+
P_{k}(t)P_{0,2}(k)\lbrack (1-\delta _{kn-1})(1-\delta
_{kn})\rbrack +
\]
\[
+P_{k}(t)P_{2,0}(k)\lbrack (1-\delta_{k1})(1-\delta _{k0}) \rbrack
\]
Substituting the values of probabilities $P_{2,0}(k)$,
$P_{0,2}(k)$, $P_{1,1}(k)$ we have
$$
P_{k}(t+1)= \frac{2(k-1)(n-k+1)\lambda}{n(n-1)}P_{k-1}(t)
(1-\delta_{k0})+
$$
\begin{eqnarray}
+\frac{2(k+1)(n-k-1)\mu}{n(n-1)}P_{k+1}(t)(1-\delta_{kn})+
\label{meq1}
\end{eqnarray}
$$
+\frac{(n-k)(n-k-1)+k(k-1)}{n(n-1)}P_{k}(t).
$$

By similar arguments one obtains the following master equations
for the MB:
$$
P_0(t+1)= \mu P_{1}(t)+  P_{0}(t),
$$
\begin{eqnarray}
P_k(t+1)=\lambda P_{k-1}(t)(1-\delta_{k,1}) + \mu
P_{k+1}(t)(1-\delta_{k,n-1}), \ \mbox{for}\ 0 < k <n, \label{meq2}
\end{eqnarray}
$$
P_n(t+1)=\lambda P_{n-1}(t) +  P_{n}(t).
$$

The equations (\ref{meq1}),(\ref{meq2}) can be used as background
of analytical
investigations of  MA and MB. We demonstrate how they allow one
to exactly calculate the important characteristics of model's dynamics.

\section{Exact results for MA }

The simplest problem for MA could be to find the stationary
solution of the master equation (\ref{meq1}).
In vector writing it looks like:
\begin{eqnarray}
P(t)-P(t-1)=-\frac{2}{n(n-1)} AVP(t-1)\ \ \ \mbox{for}\ t>0  \label{era1}
\end{eqnarray}
where $P(t)$ is the vector with components $\{P(t)\}_k=P_k(t)$,
$k=0,1,...,n$,  and $A$,$V$ are the matrices:
\[
V_{ij} =\delta _{ij}i(n-i) ,\ \ A_{ij} = \delta _{ij} -\lambda
\delta_{ij+1} -\mu \delta_{ij-1}
\]
The stationary solution $P(t)=P$ of (\ref{era1})  satisfies the
equation:
\begin{eqnarray}
AVP=0  \label{era2}
\end{eqnarray}
Since $\det A \neq 0$, it follows from (\ref{era2}) that $VP=0$.
Hence, the solution of (\ref{era2}) can be written as:
\begin{eqnarray}
\{P\}_i =\delta_{i0}\rho +\delta_{in}(1-\rho) ,\ \ 0\leq
\rho \leq 1 \label{era3}
\end{eqnarray}

Now, the problem is to find $\rho $ as the function of initial
probability distribution $P_k(0)\equiv \{P_0\}_k$. If we denote
\[
(I-\frac{2}{n(n-1)}AV)=G
\]
then in virtue of (\ref{era1}), the vector $P$ is expressed
through the vector $P_0$ in the following way
\begin{eqnarray}
P= G_{as} P_0, \ \mbox{where}\  \
G_{as}=\lim_{t\rightarrow\infty}G^t. \label{era4}
\end{eqnarray}
It follows  from (\ref{era3}),(\ref{era4}) that the matrix
$G_{as}$ must be of the form
\[
\{G_{as}\}_{ik}=\delta _{i0}x_{k}+\delta _{in}y_{k}
\]
The matrix G has the property:
\[
\sum_{i=0}^{n}\{G\}_{ik}=1\ \ \ \ \ \ \ \mbox{for}\ \ \ \ \ \
0\leq k\leq n,
\]
and  the same equality must be fulfilled for $G_{as}$ too, hence
$x_k+y_k=1$.
Therefore
\[
\{G_{as}\}_{ik}=\delta _{i0}x_{k}+\delta _{in}(1-x_{k}).
\]
The vectors $P^{(0)}$, $P^{(n})$ with components $P^{(0)}_k=\delta_{0k}$,
$P^{(n)}_k=\delta_{nk}$ are the eigen ones for matrix $G$:
$GP^{(0)}=P^{(0)}$, $GP^{(n)}=P^{(n)}$. Hence,
$G_{as}P^{(0)}=P^{(0)}$, $G_{as}P^{(n)}=P^{(n)}$, and
\begin{eqnarray}
x_{0}=1,\ \ \ x_{n}=0.
\label{era4b}
\end{eqnarray}
Taking into account that
\[
G_{as}G=G_{as},
\]
we obtain:
\[
x_{l}G_{lk}=x_{k}.
\]
Thus, we have to solve the equation
$
xAV=0
$
which can be written in components as
\begin{eqnarray}
x_{j}-\lambda x_{j+1}-\mu{x_{j-1}}=0,\ \ \ \ 0<\ j<n \label{era5}
\end{eqnarray}
The equations (\ref{era5})
with boundary condition (\ref{era4b})
coincide with ones of the classical problem of the player's losing
\footnote {It was pointed out us by A.Capocci.}(see for example eq.
(2.1),(2.2) of capital XIV in \cite{feller}).
The  solution
of (\ref{era4b}), (\ref{era5}) has the form
\[
x_{k}=\mu^{k}\frac{\lambda^{n-k}-\mu^{n-k}}{\lambda^{n}-
\mu^{n}} = \frac{\omega^{k}-\omega ^{n}}{1-\omega^{n}}.
\]
Here, we used the convenient notation $\omega=\mu/\lambda$.

In virtue of (\ref{era3}),(\ref{era4}), the parameter $\alpha$
defining the stationary solution of the master equation can be
expressed in terms of initial probability distribution $P_k(0)$ as
follows
\begin{equation}
\rho = \sum_{k=0}^{n}P_{k}(0)x_{k}=\frac{ P\big(\omega\big)-\omega
^{n}}{1-\omega ^{n}}
\label{rho}
\end{equation}
We denoted $P(\omega)$ the generating function
\[
P(\omega)=\sum_{k=0}^{n} P_{k}(0)\omega^{k}.
\]
For the homogeneous initial distribution $P_{k}(0)=\frac{1}{n+1}$
we have:
\[
P(\omega)=\frac{1}{n+1}\frac{1-\omega^{n+1}}{1-\omega}, \ \
\rho=\frac{1-(1+n)\omega^{n}+n\omega^{n+1}} {(n+1)(1-\omega
)(1-\omega^{n})}.
\]
For large $n$ we get
\[
\rho=1-\frac{\omega}{(\omega-1)(n+1)}+\omega^{-n}nQ_1(n), \ \
\mbox{if}\ \ \omega > 1,
\]
\[
\rho =\frac{1}{(n+1)(1-\omega)}+\omega^{n}Q_2(n), \ \ \mbox{if}\
\ \omega < 1,
\]
\[
\rho =\frac{1+e^{x}(x-1)}{x(e^{x}-1)}+ \frac{Q_3(n)}{n}, \ \
\mbox{if}\ \ \omega=1+\frac{x}{n}.
\]
Here, the function $Q_1(n)$, $Q_2(n)$,  $Q_3(n)$ have finite
limits for $n\rightarrow \infty$.

\section{Exact solutions for MB}

Now, we consider the master equations for MB.
For the system with elements of qualities $p$ and $q$  this model
can be considered as a reformulated  classical player's losing model
\cite{feller}. Let us denote $P(z,u)$ the generating function of
probability distribution $P_k(t)$:
\[
P(z,u)=\sum_{k=0}^{n}\sum_{t=0}^{\infty} P_{k}(t)z^{k}u^t,
\]
and will use the notations
\begin{eqnarray}
A(u)=P(0,u)=\sum_{t=0}^{\infty}P_0(t)t^u,\ \
B(u)=\frac{1}{n!}\frac{\partial^n}{\partial
z^n}P(z,u)\Bigg|_{z=0} =  \sum_{t=0}^{\infty}P_n(t)t^u.
\label{exmb}
\end{eqnarray}
Then the equations (\ref{meq2}) can be rewritten for generating
function as
$$
\frac{P(z,u)-P(z,0)}{u}=\left(\lambda z +\frac{\mu}{z}\right)P(z,u)+
\left(1-\lambda z -\frac{\mu}{z}\right)[A(u)+z^n B(u)]
$$
or in an equivalent form:
\begin{eqnarray}
P(z,u)[z-u(\lambda z^2 +\mu)] =
z P(z,0)+u\left(z-\lambda z^2-\mu\right)[A(u)+ z^{n}B(u)]
\label{esb1}
\end{eqnarray}
If the functions $A(u)$, $B(u)$ are  known, the solution of
equation (\ref{esb1}) is the following
$$
P(z,u) =\frac{
z P(z,0)+u\left(z-\lambda z^2-\mu\right)[A(u)+ z^{n}B(u)]}
{z-u(\lambda z^2 +\mu)}
$$

Let us denote $\alpha_1(u)$, $\alpha_2(u)$
the solutions of
equation
$
z-u(\lambda z^2 +\mu) = 0
$:
$$
\alpha_1(u)\equiv\frac{1-\sqrt{1-4u^2\lambda\mu}}{2u\lambda},\ \
\alpha_2(u)\equiv\frac{1+\sqrt{1-4u^2\lambda\mu}}{2u\lambda},
$$
then  substituting $z=\alpha_1=\alpha_1(u)$ and $z=\alpha_2=\alpha_2(u)$
in (\ref{esb1}) we obtain
two equations for $ A=A(u)$, $B=B(u)$
\begin{eqnarray}
A+\alpha_1^nB=\frac{\alpha_1P_1}
{u(\lambda \alpha_1^2 -\alpha_1 +\mu)},\ \
A+\alpha_2^nB=\frac{\alpha_2P_2}
{u(\lambda \alpha_2^2 -\alpha_2+\mu)}
\label{esb2}
\end{eqnarray}
Here we denoted $P_i=P(\alpha_i(u),0)$, $i=1,2$.
The solution of the equations (\ref{esb2}) has the form:
\begin{eqnarray}
\label{esb3}
A(u)
=\frac{\alpha_2^n P_1-\alpha_1^n P_2}
{(1-u)(\alpha_2^n-\alpha_1^n)}
,\ \
B(u)
=\frac{ P_1- P_2}
{(1-u)(\alpha_1^n-\alpha_2^n)}
\end{eqnarray}

Thus, we have obtained
\begin{eqnarray}
P(z,u) =\frac{ zP(z,0)}
{[z-u(\lambda z^2 +\mu)]}+
\label{solmb}
\\
+
\frac{u\left(z-\lambda z^2-\mu\right)
[(z^n-\alpha_2^n)P_1-(z^n-\alpha_1^n)P_2]}
{(1-u)[z-u(\lambda z^2 +\mu)](\alpha_1^n-\alpha_2^n)}
\nonumber
\end{eqnarray}
It is the presentation in terms of generating function of exact
solution for the master equation (\ref{meq2}) for the finite
system with $n$ elements.

For the limit of infinite system $n\rightarrow\infty$ we obtain from
(\ref{esb1})
more simple equation:
\begin{eqnarray}
P(z,u)[z-u(\lambda z^2 +\mu)] =
z P(z,0)+u\left(z-\lambda z^2-\mu\right)A(u).
\label{esb4}
\end{eqnarray}
It follows from (\ref{esb4}) that
$$
A(u)=\frac{P(\alpha(u),0)}{1-u},
$$
where $\alpha(u)=\alpha_1(u)$ is the analytical
at the point $u=0$
solution
of equation
$
z-u(\lambda z^2 +\mu) = 0
$.
Thus, the  solution of (\ref{esb4}) has the form:
\begin{eqnarray}
P(z,u) =\frac{
z (1-u)P(z,0)+u\left(z-\lambda z^2-\mu\right)P(\alpha(u),0}
{(1-u)[z-u(\lambda z^2 +\mu)]}
\label{esb5}
\end{eqnarray}

The moments $M^{(n)}(t)$ of the considered distribution function
$P_n(t)$ defined as
$$
M^{(s)}(t)=\sum_{k=0}^{n}P_k(t)k^s,
$$
can be found by differentiating the generating function $P(z,u)$.
Particularly,
$$
M^{(1)}(t)=\frac{1}{t!}
\frac{\partial}{\partial z}\frac{\partial^t}{\partial u^t}
P(z,u)\Bigg|_{z=1,u=0},
$$
$$
M^{(2)}(t)=\frac{1}{t!}
\frac{\partial^2}{\partial z^2}\frac{\partial^t}{\partial u^t}
P(z,u)\Bigg|_{z=1,u=0} +M^{1}(t)
$$
Hence for $M^1(t)$, $M^2(t)$ can be expressed in
terms of power series
of coefficients of the functions
$$
m_1(u)=\frac{\partial}{\partial z}P(z,u)\Bigg|_{z=1}, \ \ m_2(u)=
\frac{\partial^2}{\partial z^2}P(z,u)\Bigg|_{z=1}
$$
Since
$$
P(1,0)=1, \ \
\frac{\partial}{\partial z}P(z,0)\Bigg|_{z=1}=M^{(1)}(0),
$$
$$
\frac{\partial^2}{\partial z^2}P(z,0)\Bigg|_{z=1}=M^{(2)}(0)+ M^{(1)}(0).
$$
It follows from (\ref{esb5}) that for infinite system
$$
  m_1(u)=\frac{M^{(1)}(0)}{1-u}+
  \frac{(2\lambda-1)u(1-P(\alpha,0))}{(1-u)^2},
$$
$$
 m_2(u)= \frac{M^{(2)}(0)}{1-u}+\frac{2(2\lambda-1)uM^{(1)}(0)}{(1-u)^2}+
\frac{2u(1+4\lambda^2 u -\lambda (1+3u))(1-P(\alpha,0))}{(1-u)^3},
$$
and
$$
M^{(1)}(t)= M^{(1)}(0)+
\frac{1-\lambda}{\lambda}
\chi_1
+(2\lambda-1)\left[1-\chi_0\right]t + \frac{[4\lambda(1-\lambda)]^
{\frac{t}{2}}}{t^{\frac{3}{2}}}{\cal O}_1(t)),
$$
$$
M^{(2)}(t)=
M^{(2)}(0)-\frac{1-\lambda}{\lambda}\chi_1 -
\frac{(1-\lambda)^2}{\lambda^2}\chi_2
+
$$
$$
+
\left[2(2\lambda-1)\left(M^{(1)}(0)
+\frac{(1-\lambda)}{\lambda}
\chi_1\right)
-4\lambda^2
(1-\chi_0)
\right]t+
$$
$$
+ (1-2\lambda)^2(1-\chi_0)t^2 +\frac {
[4\lambda(1-\lambda)]^{\frac{t}{2}} } { t^{\frac{3}{2}} } {\cal
O}_2(t)
$$
Here ${\cal O}_i(t) =0$, $i= 1,2$ are limited in the region of
large $t$ functions:
$$
|{\cal O}_i(t)|< {\cal C }\ \ \ \mbox{for} \  t >> 1,
$$
where ${\cal C} $ is a constant, and
$$
\chi_0=P(\omega,0),\ \
\chi_1=\frac{d}{dx}P(x,0)|_{x=\omega}, \ \
\chi_2=\frac{d^2}{dx^2}P(x,0)|_{x=\omega}.
$$

We call filtering to be finished at time point $t$ if
at this moment the quality of all the element became first to be
equal. Let us denote $P^{(p)}(t)$, $P^{(q)}(t)$  the probabilities
that filtering is finished at time point $t$ with quality of all the
element being equal to $p$ and $q$ consequently.
We have
$$
 P_0(t+1)=P^{(p)}(t+1)+P_0(t),\ \ P_n(t+1)=P^{(q)}(t+1)+P_n(t).
$$
These relations can be rewritten on terms of generating
functions as
$$
{\cal P}^{(p)}(u) = A(u)(1-u),\ \ {\cal P}^{(q)}(u)=B(u)(1-u).
$$
where $A(u)$, $B(u)$ are defined in (\ref{exmb}) and
${\cal P}^{(p)}(u)$ (${\cal P}^{(p)}(u)$) is the generating
function of probabilities $P^{(p)}(t)$ ($P^{(p)}(t)$):
$$
{\cal P}^{(p)}(u)=\sum_{t=0}^\infty P^{(p)}(t)u^t, \ \
{\cal P}^{(q)}(u)=\sum_{t=0}^\infty P^{(q)}(t)u^t.
$$
In virtue of (\ref{esb3}), it follows that
\begin{equation}\label{ft}
{\cal P}^{(p)}(u)=
\frac{
\alpha_1^n P_2 -\alpha_2^n P_1}
{\alpha_1^n-\alpha_2^n}, \ \ {\cal P}^{(q)}(u)=
\frac{
P_1 - P_2}
{\alpha_1^n-\alpha_2^n}
\end{equation}
Hence, for generating  function ${\cal P}(u)$ of probability
$P(t)=P^{(p)}(t)+P^{(q)}(t)$ that filtering is finished at the moment
$t$ (called in \cite{Zhang} search time) we obtain:
\begin{equation}\label{ft}
{\cal P}(u)\equiv \sum_{t=0}^\infty P(t)u^t=
\frac{
(1 -\alpha_2^n) P_1-(1-\alpha_1^n) P_2}
{\alpha_1^n-\alpha_2^n}
\end{equation}
For the mean value $T$ of the filtering (search) time  we obtain:
\begin{equation}\label{mft}
T= \frac{\partial}{\partial u}{\cal P}(u)|_{u=1}=
\frac{M^{(1)}(0)}{\mu-\lambda}-n\frac{1-P(\omega )}{(\mu-\lambda)
(1-\omega^{n})}
\end{equation}
This result agrees with known formula for mean time of play in the
problem of player's losing (see for example capital XIV,
formula (3.5) in \cite{feller}). For the homogeneous initial probability
distribution $P_k(0)=1/(n+1)$ we have:
\begin{equation}\label{mft}
T=
n\frac{
(1+\omega)(1-\omega^{n})-
n(1-\omega)(1+\omega^{n})}
{2(n+1)(\mu-\lambda)(1-\omega)
(1-\omega^{n})},
\end{equation}
and for large $n$: $T=n/2|\mu-\nu|+{\cal O}(n)$, $|{\cal O}(n)|< C$,
where $C$ is a constant.

The stationary solution of the master equation for the MB can be found
as the residue of the pole in the point $u=1$ of $P(z,u)$. For the
generating function $P_{st}(z)=\sum_{k=0}^n P_k z^k $ of stationary
distribution $P_k$, we obtain from (\ref{solmb}) the following
result
\begin{equation}
P_{st}(z)= \mbox{res}_{u=1}[P(z,u)]= \frac{
(z^n-1)P(\omega)-(z^n-\omega^n)}
{\omega^n-1}.
\label{mbst}
\end{equation}
Comparing (\ref{era3}),(\ref{rho}) and (\ref{mbst}), we see that
the stationary solutions of the master equations for MA and MB coincide.

\section{Conclusion}

We studied simple processes of information filtering generated
by consequent comparisons of two randomly chosen elementary
information units.  For the simplest version of MB the mean search
time $T$ is given by (\ref{mft}). It follows directly from
dynamical rules that the search time for MA must be larger.
The search time is dependent on initial distribution too.
Nevertheless, the obtained numerical and analytical results
shows that for large system the search time $T$ and the number
$N$ of the elements obey the relation $T/N=C$, where $C$ is
independent on $T$ and $N$ and depend on  initial
distribution and qualities of elements only. For the MB with
elements of quality $p$, $q$ and  homogeneous initial distribution
$P_k(0)=1/(n+1)$, quantity $C$ looks like $C=(p+q)/2|p-q|$ and
and becomes large for small $|p-q|$.

The introduced in \cite{Zhang} characteristic
of information filtering called efficiency $R$ is defined as the
rank of selected value in the starting configuration.
The efficiency for the MA and MB with elements of the quality
$p$ and $q$ can be estimated by using the stationary solution
of the master equations (\ref{era3}). If $p>q$ the quantity
${\cal R}=(1-\rho)/\rho $ is the ratio of probabilities to select
the elements with qualities $p$ and $q$. It can be considered
as a measure of filtration efficiency. For homogeneous initial
distribution and large number $N$ of the system elements
${\cal R} = N(1-\omega)$.

The information filter investigated in \cite{Zhang} is characterized
for large systems by search time $T\sim N^2$ and efficiency
$R \sim \ln N$.
Comparison these results with our ones shows
that the information filtration  respecting
one dimensional organization of information units
appears to be slower and less effective as
one based on the random choosing algorithm.

\vspace {0,5cm} \centerline{\bf Acknowledgments} \vspace {0,5cm}

The authors would like to thank Andrea Capocci and Yi-Cheng Zhang
for many fruitful  discussion.
Yu.M. Pismak was supported in part by
the Russian Foundation of Basic Research ( Grant No 00-01-00500)
and by the Schweizerische Nationalfonds (SNF) ( SCOPES Grant
No 7SUPJ062295).

\begin{figure}[t]
  \begin{center}
    \epsfig{file= 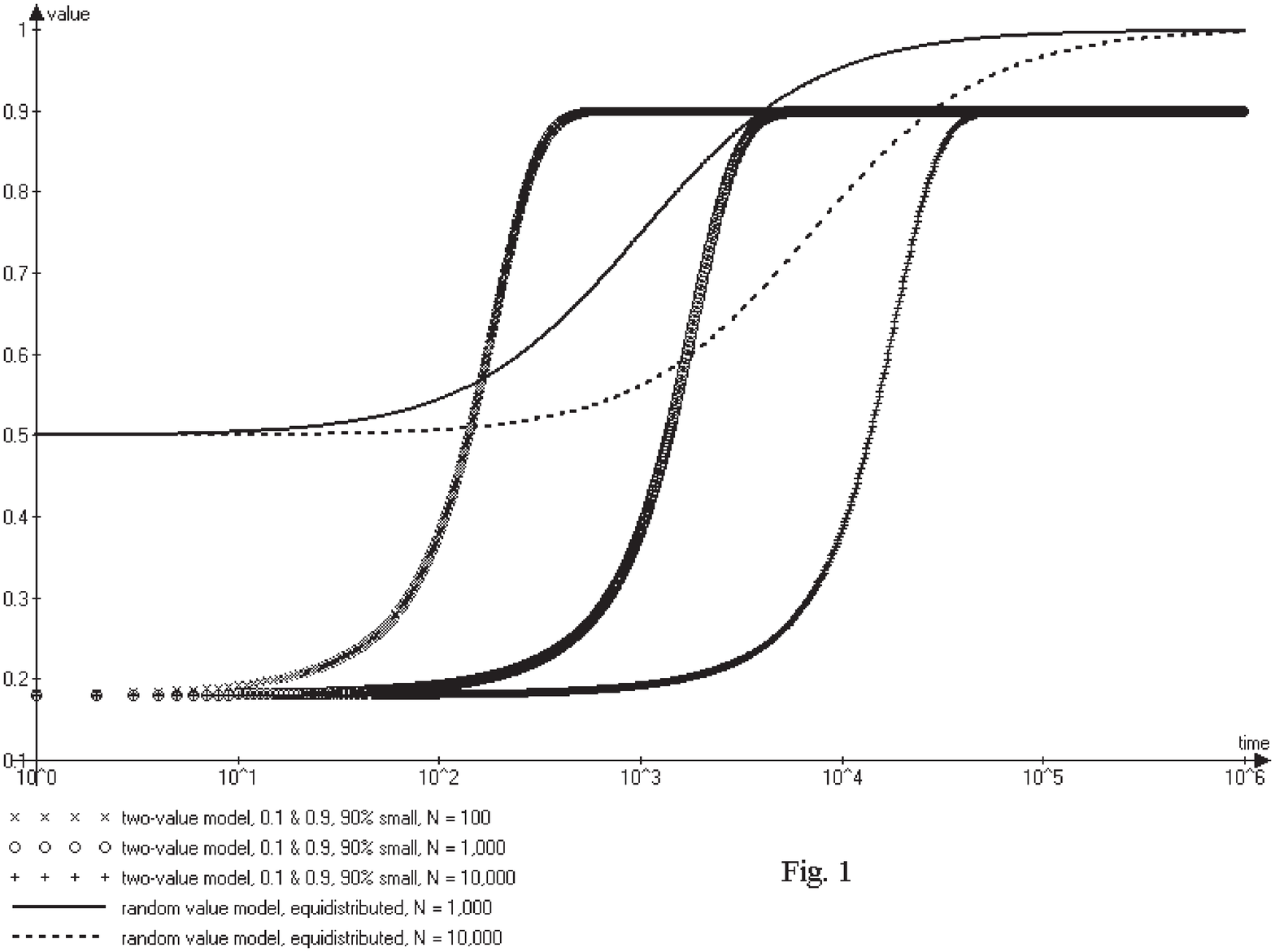,height=9cm,angle=0}
\label{label}
\end{center}
\end{figure}

\begin{figure}[t]
  \begin{center}
    \epsfig{file= 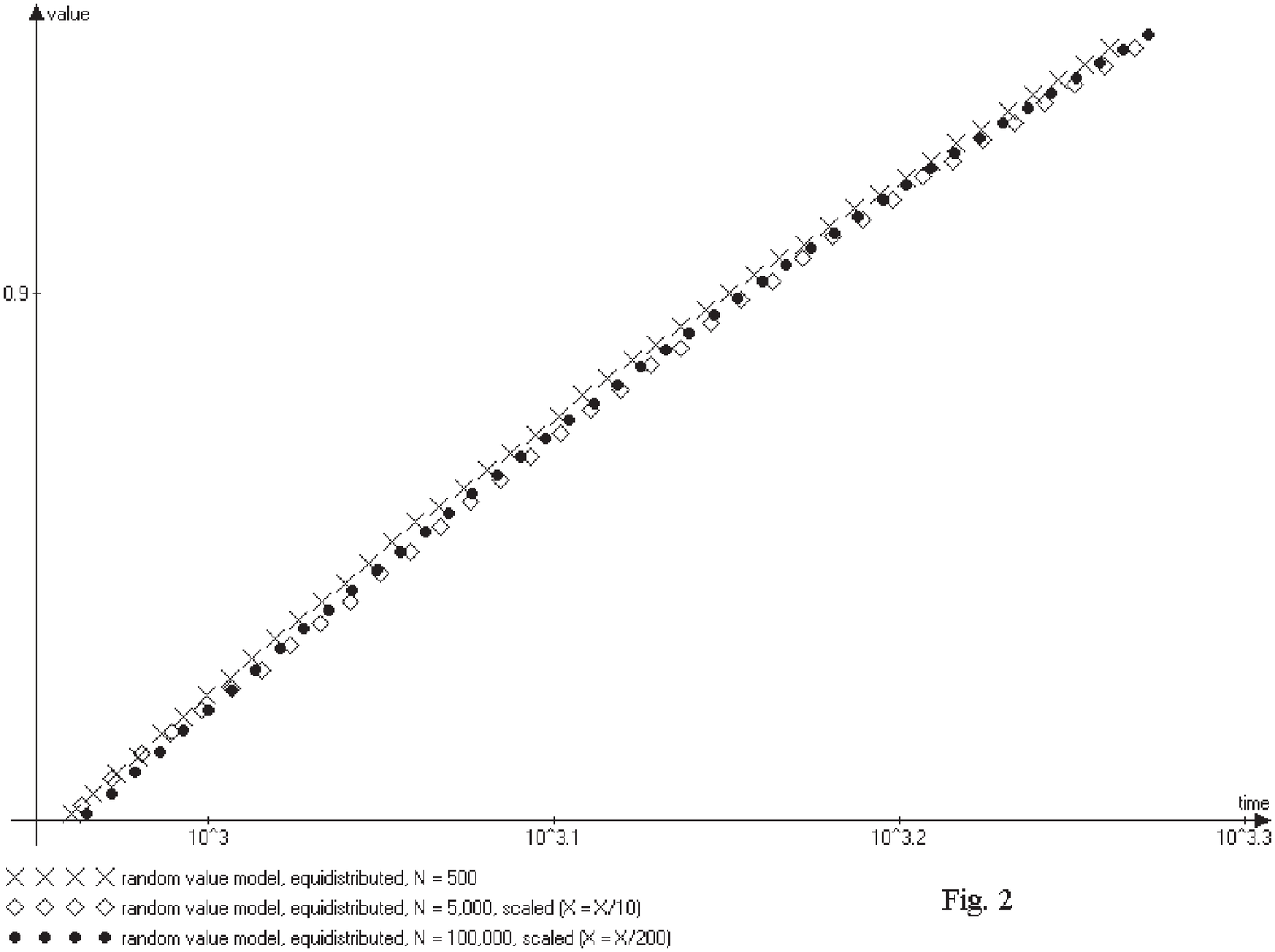,height=9cm,angle=0}
\label{label}
\end{center}
\end{figure}

\begin{figure}[t]
  \begin{center}
    \epsfig{file= 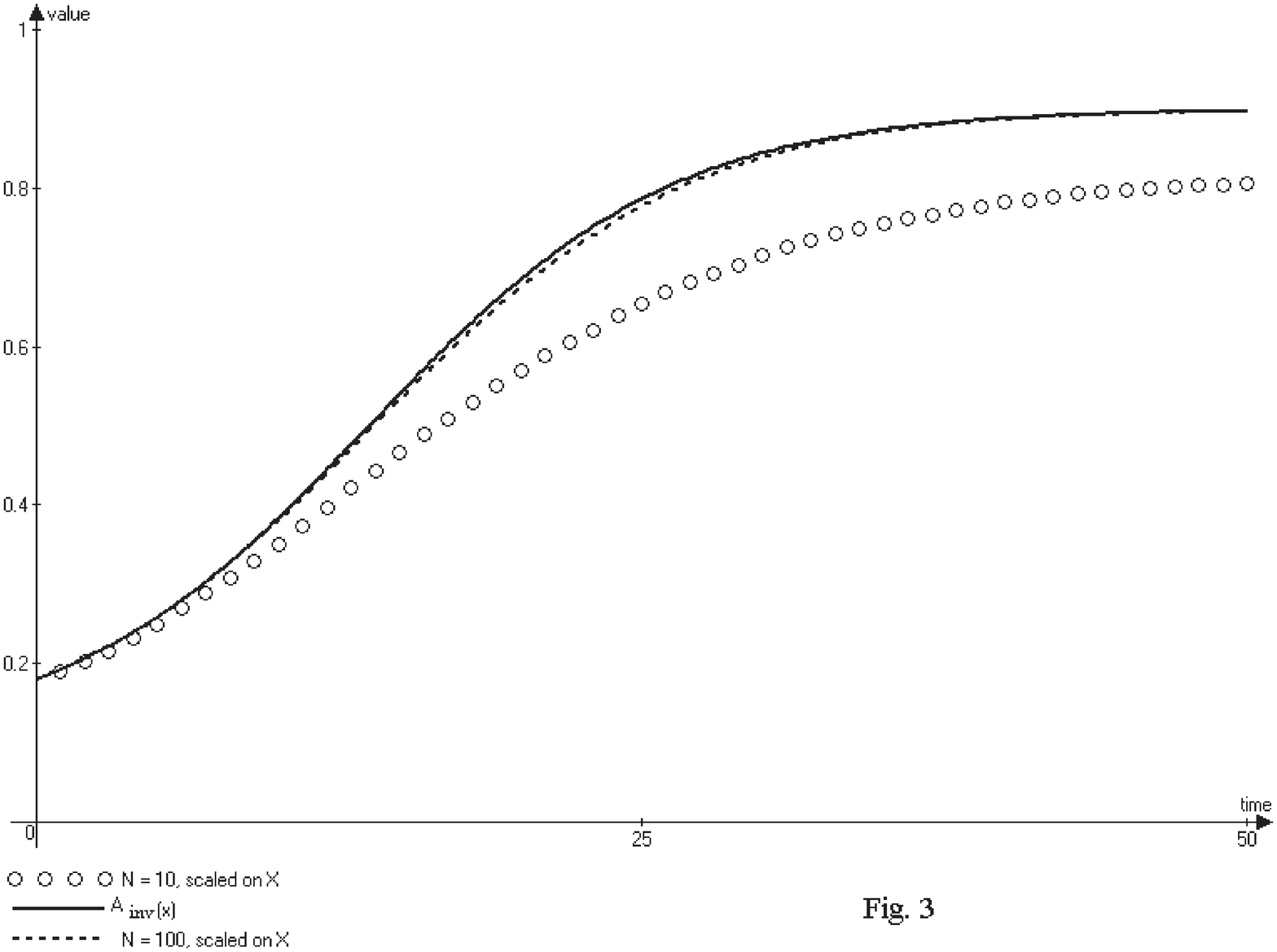,height=9cm,angle=0}
\label{label}
\end{center}
\end{figure}

\begin{figure}[t]
  \begin{center}
    \epsfig{file= 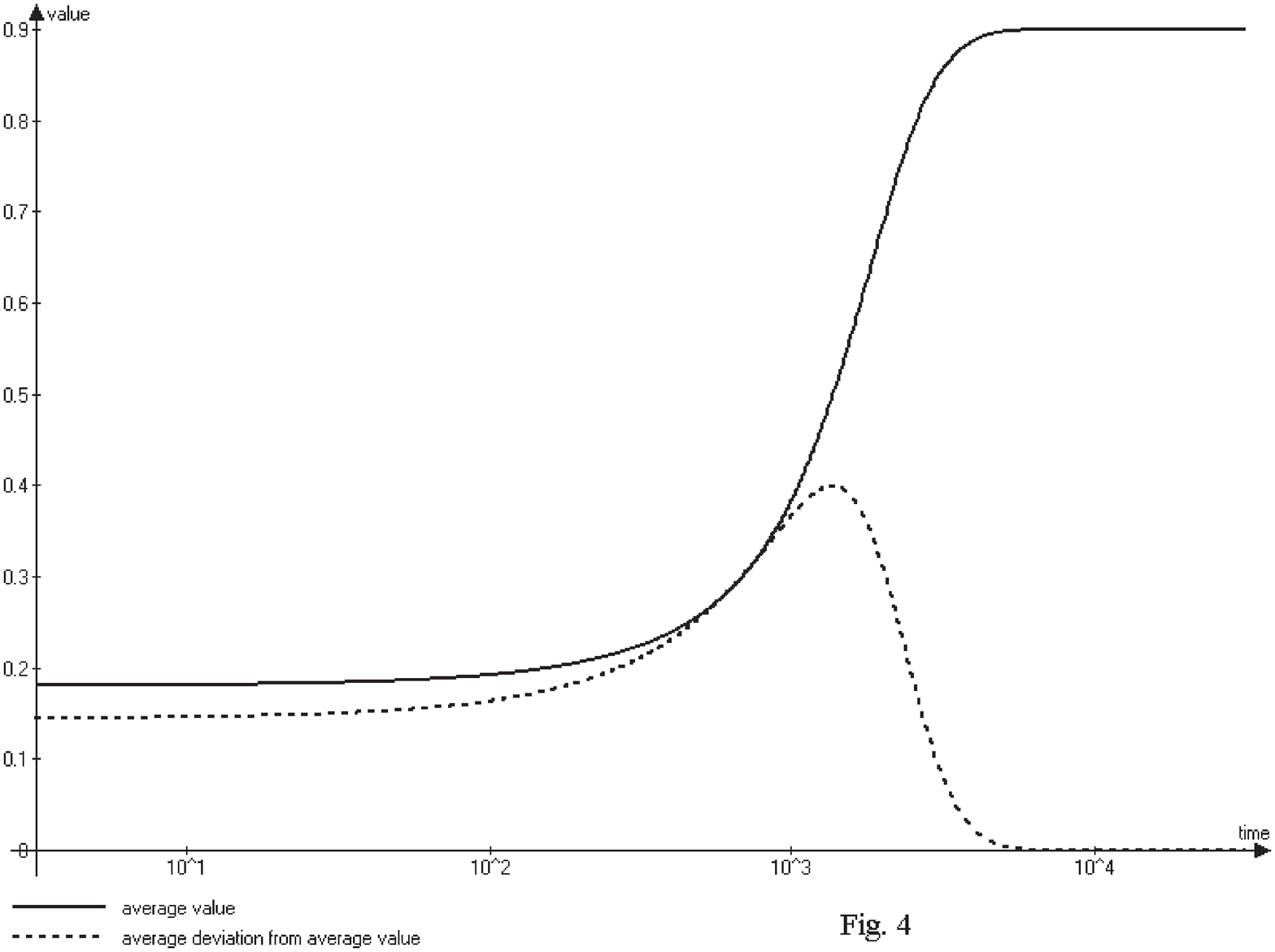,height=9cm,angle=0}
\label{label}
\end{center}
\end{figure}

\begin{figure}[t]
  \begin{center}
    \epsfig{file= 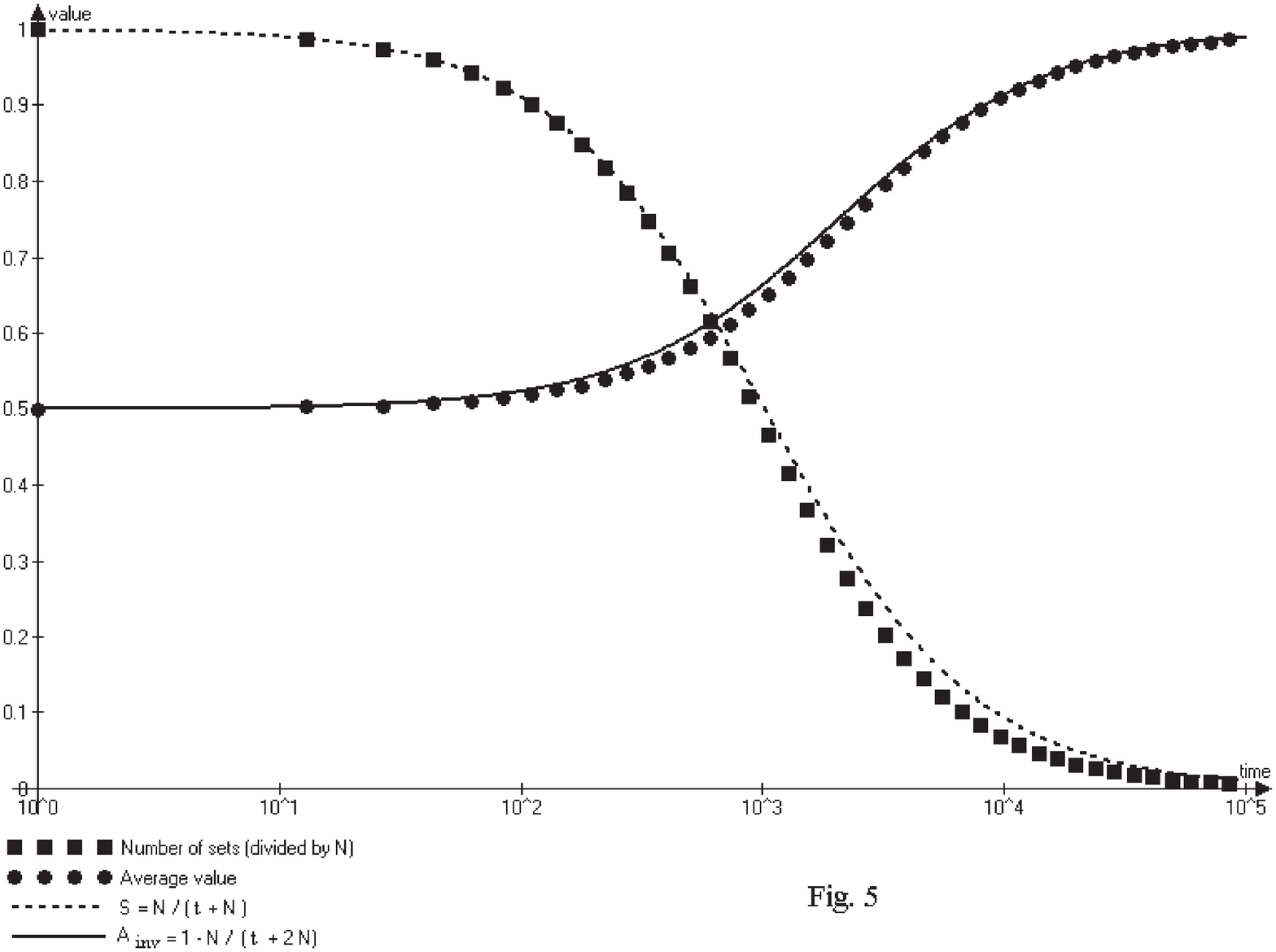,height=9cm,angle=0}
\label{label}
\end{center}
\end{figure}

\begin{figure}[t]
  \begin{center}
    \epsfig{file= 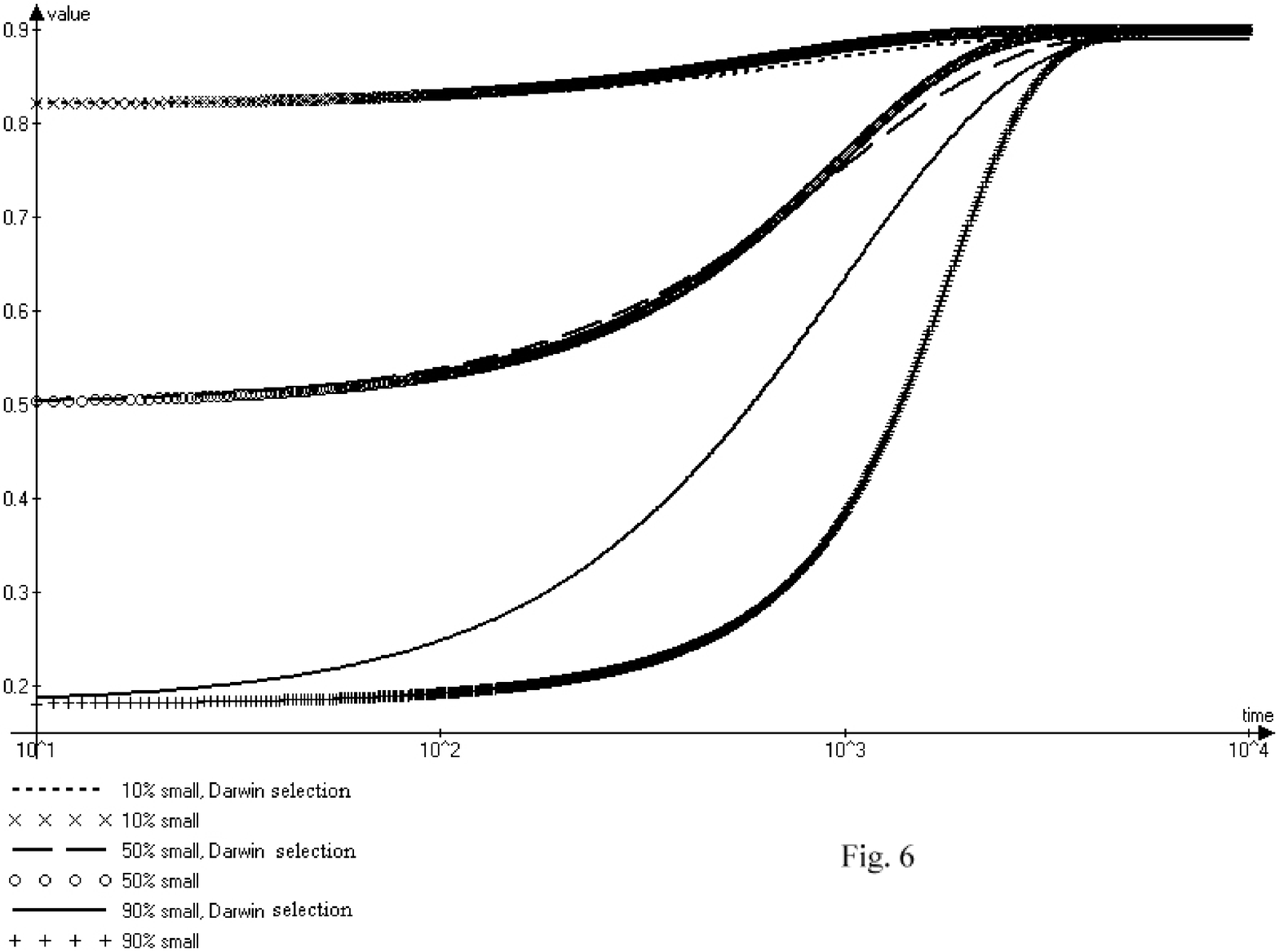,height=9cm,angle=0}
\label{label}
\end{center}
\end{figure}

\end{document}